\documentclass[a4paper,12pt]{article}
\usepackage[english]{babel}
\usepackage[utf8]{inputenc}
\usepackage{t1enc}
\usepackage{floatflt}
\usepackage{graphicx}
\usepackage{psfrag}
\usepackage{bbm}
\usepackage{amsmath}
\usepackage{amssymb}
\usepackage{hyperref}
\usepackage{ifthen}
\usepackage{subcaption}
\usepackage{epstopdf}

\def\d{\mathrm{d}}
\def\e{\mathrm{e}}
\def\imagi{\mathrm{i}}

\def\lag{{\mathcal{L}}}

\def\kihagy#1{}

\newcommand{\doix}[2]{\href{http://dx.doi.org/#2}{#1}}%
\newcommand{\arxiv}[2][]{%
  \ifthenelse{\equal{#1}{}}{%
    \href{http://arxiv.org/abs/#2}{\texttt{arXiv:#2}}%
  }{%
    \href{http://arxiv.org/abs/#2}{\texttt{arXiv:#2 [#1]}}%
  }%
}%

\title{Charge screening in the Abelian Higgs model}
\author{Péter Forgács\textsuperscript{1,2} and Árpád Lukács\textsuperscript{3,1}\\
{\small {}\textsuperscript{1} Wigner RCP RMI, H1525 Budapest, POB 49.}\\
{\small {}\textsuperscript{2} Institut Denis-Poisson CNRS/UMR 7013, Universit\'e de Tours,}\\
{\small Parc de Grandmont, 37200 Tours, France}\\
{\small {}\textsuperscript{3} Department of Theoretical Physics, University of the Basque Country UPV/EHU,}\\
{\small POB 644, E-48080 Bilbao, Spain}
}

\begin{document}

\maketitle

\begin{abstract}
In the Abelian Higgs model electric (and magnetic) fields of external charges (and currents) are screened by the scalar field. In this contribution, complementing recent investigations of Ishihara and Ogawa, we present a detailed investigation of charge screening using a perturbative approach with the charge strength as an expansion parameter. It is shown how  perfect global and remarkably good local screening can be derived from Gauss' theorem, and the asymptotic form of the fields far from the sources. The perturbative results are shown to compare favourably to the numerical ones.
\end{abstract}

Historically, the idea for a renormalisable model containing massive gauge fields (of which the Abelian Higgs model is arguably the simplest one) came from the theory of shielding of electromagnetism in a superconductor \cite{anderson, higgs}.

The analogy between how the Abelian Higgs model converts a long-range interaction (mediated by a massless gauge field) into a short-range one (mediated by a massive vector field) and how electromagnetism is shielded by mobile charges is well-known. Motivated by the recent investigation of charge screening in the Abelian Higgs model in Ref.\ \cite{ishiharaogawascreen}, and two subsequent papers, Refs.\ \cite{ishiharaogawasoln, ishiharaogawasoln2}, where the results are applied, we present a simple, systematic perturbative framework to study this phenomenon analytically, complementing those studies. One is quickly led to believing that all this must have long been done, since, e.g., screening in scalar electrodynamics has been considered in Refs.\ \cite{mandula, adlerpiran, bawincugnon}, and in Yang-Mills theories in Refs.\ \cite{sikivieweiss, campbellnorton}. To the best of our knowledge, however, prior to Ref.\ \cite{ishiharaogawascreen} there has been no detailed investigation published about this ``schoolbook'' case.

This note is intended as a (pedagogical) complement of Ref.\ \cite{ishiharaogawascreen} where the physics of screening in the Abelian Higgs model has been studied numerically and also analytically by various approximations for both point-like and for extended charge distributions.
Our note is tuned to a perturbative study of the physics of charge screening, considering the external charge strength as a ``small'' parameter and compute corrections to the usual massive Green's function. In this approach it is easy to investigate local charge screening, while global screening is shown to follow from Gauss' theorem \cite{FLQ}. The perturbative results are shown to agree remarkably well with the numerical ones.

The structure of the paper is as follows: in Sec.\ \ref{sec:model} we introduce the notations used in the model, then in Sec.\ \ref{sec:gauge} choose a gauge and a suitable set of variables. Sec.\ \ref{sec:static} contains the field equations for static solutions, it is exhibited how perfect screening follows from the asymptotic behaviour of the fields and Gauss' theorem and the setup of a series solution. In Sec.\ \ref{sec:ptsrc} we consider the field of a point source, using both perturbation theory and numerical methods. In Sec.\ \ref{sec:Gauss}, we calculate the field of a Gaussian charge distribution, in Sec.\ \ref{sec:hom}, that of a homogeneously charged sphere, and then conclude. Appendix\ \ref{app:Yukawa}  contains the derivation of the Green function in the static case. Appendix\ \ref{app:KGGreen} is devoted to present the Green function of the time-dependent massive Klein-Gordon equation in a closed form in coordinate space representation, and demonstrate its consistency with the Green function of the static equation. In Appendices\ \ref{app:KGInt} resp.\ \ref{app:PInt} the interaction potential between two point sources in a Klein-Gordon resp.\  a Proca field are calculated (using the method of Refs.\ \cite{LL2, speight}).

\section{The model considered}\label{sec:model}
The Abelian Higgs model contains the following fields: a $U(1)$ gauge field $A_\mu$, whose mass is to be generated by the Higgs field, a complex scalar field $\phi$, which is subject to self-interaction due to a potential, whose minimum is obtained at a non-zero value of $\phi$. The model is specified by the action integral,
\begin{equation}\label{eq:action}
 S = \int \d^4 x \lag\,,\quad \lag = -\frac{1}{4}F_{\mu\nu}F^{\mu\nu} + (D_\mu\phi)^* D^\mu \phi - V(\phi^*,\phi) -A_\mu j^\mu_{\rm ext}\,,
\end{equation}
where there is an implicit summation over repeated Greek indices from 0 to 4, indices are moved up and down with the help of the Minkowski metric, $g_{\mu\nu} = {\rm diag}(1,-1,-1,-1)$ (sign convention of Ref.\ \cite{LL2}), Greek indices run from 0 to 3, and Latin (spatial) ones from 1 to 3. In this paper, we shall use units where the unit of length and time agree (velocity of light is unity).

The field strength tensor $F_{\mu\nu}=\partial_\mu A_\nu - \partial_\nu A_\mu$ is antisymmetric, containing the electric and magnetic fields as $F_{0i} = E_i$, $F_{ij} = \varepsilon_{ijk}B_k$, where $\epsilon_{ijk}$ is the fully antisymmetric unit tensor in flat 3d space, $\varepsilon_{123} = 1$. The gauge covariant derivative of the scalar field is given as $D_\mu \phi = (\partial_\mu -\imagi e A_\mu)\phi$, and the potential is
\[
 V(\phi,\phi^*) = \frac{\lambda}{4}(\phi^*\phi-\eta^2)^2\,,
\]
where $e$ is the electric charge and $\eta$ a constant.

The vector field $j^\mu_{\rm ext}$ shall always be a fixed (external) current density, assumed to be conserved, $\partial_\mu j^\mu_{\rm ext} =0$. The induced current, 
\begin{equation}
 j^\mu_\phi = \imagi e\left[\phi (D^\mu \phi)^* - \phi^* D^\mu\phi\right]\,,
\end{equation}
is also the Noether current corresponding to the symmetry $\phi\to \e^{\imagi e\alpha}\phi$, $\alpha$ a real parameter ($U(1)$ phase symmetry), therefore it is conserved, $\partial_\mu j^\mu_\phi =0$.

The Euler-Lagrange equations corresponding to the action (\ref{eq:action}) are the Maxwell equations for the gauge field $A_\mu$, with the source being the sum of the external and the induced currents,
\begin{equation}\label{eq:Maxwell}
 \partial_\nu F^{\mu\nu} = -j^\mu\,,\quad j^\mu = j^\mu_\phi + j^\mu_{\rm ext}\,,
\end{equation}
and the scalar field equation
\begin{equation}\label{eq:scalar}
 D_\mu D^\mu \phi = -\frac{\partial V}{\partial \phi^*} = -\frac{\lambda}{2}(\phi^*\phi-\eta^2)\phi\,.
\end{equation}

\section{Gauge choice and identification of degrees of freedom}\label{sec:gauge}

The action (\ref{eq:action}) is invariant to transformations of the kind
\begin{equation}\label{eq:gaugetrf}
 \phi \to \phi' = \e^{\imagi e \xi}\phi\,,\quad A_\mu \to A_\mu' = A_\mu + \partial_\mu \xi\,,
\end{equation}
where $\xi$ is an arbitrary smooth function of space and time. In the Abelian Higgs model, a customary gauge fixing is the \emph{unitary gauge}, where the complex scalar field $\phi$ becomes real. Parametrising $\phi$ as
\begin{equation}\label{eq:param}
 \phi = \left(\eta+\frac{\chi}{\sqrt{2}}\right) \e^{\imagi \theta}\,,\quad \chi = \sqrt{2}(|\phi|-\eta)\,,\quad \theta = \mathop{\rm arg}\phi\,,
\end{equation}
the unitary gauge corresponds to the choice of a gauge function $\xi = -\theta/e$. Expressing the Lagrangian $\lag$ with the new variable $\chi$ one obtains:
\begin{equation}\label{eq:actionRho}
 \lag = -\frac{1}{4}F_{\mu\nu}F^{\mu \nu} + \frac{1}{2}m_A^2 A_\mu A^\mu
 + \frac{1}{2}\partial_\mu \chi \partial^\mu \chi -\frac{\lambda}{16}(\chi^2+2\chi v)^2
 +e^2 v A_\mu A^\mu \chi + \frac{e^2 }{2}A_\mu A^\mu \chi^2-A_\mu j^\mu_{\rm ext}\,,
\end{equation}
where $m_A^2 = 2 e^2 \eta^2$, $v=\sqrt{2}\eta$. The physical degrees of freedom of the theory are now manifest: $A_\mu$ has been turned into a massive (Proca) vector field, $\chi$ is a massive, self-interacting scalar, and there is a somewhat unusual coupling between the two.

The field equations are
\begin{equation}\label{eq:Proca}
 \partial_\nu F^{\mu\nu} - m_A^2 A^\mu = -{j'}_\phi^\mu - j^\mu_{\rm ext}\,,
\end{equation}
(known as the Proca equation), where the induced current is
\begin{equation}\label{eq:current}
 {j'}^\mu_\phi = -2 e^2 v \chi A^\mu -e^2\chi^2 A^\mu\,,
\end{equation}
and
\begin{equation}\label{eq:scalarRho}
 \partial_\mu \partial^\mu \chi = e^2 A_\mu A^\mu (\chi+v) -\frac{\lambda}{4}(2v^2+3v\chi+\chi^2)\chi\,.
\end{equation}
Taking the divergence of the Proca equation (\ref{eq:Proca}) and using the antisymmetry of the field strength tensor $\partial_\mu \partial_\nu F^{\mu\nu}=0$, one obtains
\begin{equation}\label{eq:Lorentz}
 m_A^2 \partial_\mu A^\mu = \partial_\mu j^\mu\,.
\end{equation}
Now assuming that the current is conserved, the resulting equation, $\partial_\mu A^\mu=0$ agrees formally with the Lorentz gauge condition in electrodynamics. Putting it another way: in the Abelian Higgs model, the unitary gauge in which the scalar field is real, implies also a Lorentz gauge condition on the vector potential. The Proca equation \eqref{eq:Proca} is therefore equivalent to
\begin{equation}\label{eq:Proca2}
 \partial_\nu \partial^\nu A^\mu +m_A^2 A^\mu= {j'}^\mu_\phi + j^\mu_{\rm ext}\,,\quad \partial_\mu A^\mu =0\,,
\end{equation}
i.e., the massive Klein-Gordon equation with an auxiliary condition (formally, the Lorentz gauge condition).

Finally, let us consider the splitting of the equations into three dimensional parts. The Proca electric, $E_i$, resp.\ magnetic, $B_i$, fields satisfy the Maxwell-type equations
\begin{equation}\label{eq:MaxwellDE}
  E_i = \dot{A}_i - \partial_i A_0\,,\quad \partial_i E_i + m_A^2 A_0 = \varrho_\phi + \varrho_{\rm ext}\,,
\end{equation}
\begin{equation}\label{eq:MaxwellRB}
 B_i = \varepsilon_{ijk}\partial_j A_k\,,\quad \varepsilon_{ijk}\partial_j B_k + m_A^2 A_i = -\dot{E}_i+j'_{\phi,i} + j_{{\rm ext}, i}\,,
\end{equation}
where we have introduced the notation $\varrho_\phi = {j'}_{\phi,0}$ and $\varrho_{\rm ext} = j_{{\rm ext},0}$.

The other two groups of Maxwell equations are of the usual form:
\begin{equation}\label{eq:Maxwell34}
 \varepsilon_{ijk}\partial_j E_k = \dot{B}_i\,,\quad \partial_i B_i = 0\,,
\end{equation}
making it possible to express $E_i$, resp.\ $B_i$, in terms of the potentials $A_0$ and $A_i$.
The energy density of a field configuration can be written as
\begin{equation}\label{eq:edens}
 \mathcal{E} = \frac{1}{2}\left({\bf E}^2 + {\bf B}^2\right) + \frac{m_A^2}{2}\left(A_0^2 + {\bf A}^2\right) + \frac{1}{2}\left({\dot{\chi}}^2 + \nabla\chi^2\right) + \frac{e^2}{2}\left(A_0^2+{\bf A}^2\right)\left(\chi^2+2v\chi\right) + V\,.
\end{equation}
The model may be rescaled by the following change of variables: $x^\mu \to x^\mu/(e\eta)$, $\rho\to \eta\rho$, $A^\mu \to \eta A^\mu$. The resulting Lagrangian assumes the same form as the original one [Eq.\ (\ref{eq:actionRho})], with an overall $1/e^2$ factor, and the following replacements: $e\to e_s=1$ in the covariant derivatives, $\lambda\to \beta$ in the potential, where $\beta=\lambda/e^2$. In the new units, the vector mass becomes $m_{As}=\sqrt{2}$ and the scalar one $m_{ss} = \sqrt{\beta}$, and the expectation value $v_s=\sqrt{2}$. For the integrated quantities, the replacement is $Q\to (1/e)Q_s$ and $E\to (\eta/e)E_s$, where the rescaled quantities are given by the same formulae as their unscaled counterparts, with the appropriately replaced parameters. In what follows, we shall use these units, and drop the index ``$s$''.

\section{Static solutions}\label{sec:static}
Let us now consider time-independent external sources, $j^\mu_{\rm ext}(t,x_i) = j^\mu(x_i)$ and seek time-independent solutions of the Proca equations (\ref{eq:Proca2}) which now become :
\begin{equation}\label{eq:timeindept}
 -(\nabla^2-m_A^2) A_0 = \varrho_\phi + \varrho_{\rm ext}\,,\quad -(\nabla^2-m_A^2) A_i = j'_{\phi, i} + j_{{\rm ext},i}\,,
\end{equation}
where $\nabla^2 = \partial_i \partial_i$. Eq.\ \eqref{eq:scalarRho} for the scalar field is written as
\begin{equation}\label{eq:timeindeptS}
 (\nabla^2-m_s^2)\chi = \frac{\beta}{4}(3v+\chi)\chi^2 - e^2(A_0^2-A_i^2)(v+\chi)\,,
\end{equation}
where $m_s^2 = \beta v^2/2$.

As the equations for the temporal and the spatial components of Eq.\ (\ref{eq:timeindept}) agree, it is sufficient to consider one (say, the temporal one) without loss of generality.

For later reference, let us also consider spherically symmetric solutions, where $A_0 = A_0(r)$ and $\chi=\chi(r)$. The resulting radial equations are
\begin{equation}\label{eq:radeq}
\begin{aligned}
 \frac{1}{r^2}(r^2 A_0')' &= m_A^2 A_0 + 2 e^2 A_0 v \chi + e^2 A_0 \chi^2 -\varrho_{\rm ext}\,,\\
 \frac{1}{r^2}(r^2 \chi')' &=  m_s^2 \chi + \frac{\beta}{4}(3v+\chi)\chi^2 - e^2 A_0^2 (v+\chi)\,.
\end{aligned}
\end{equation}
In Eq.\ (\ref{eq:radeq}), a prime on radial functions denotes $\d/\d r$.

\subsection{Asymptotic solution}\label{ssec:asy}
Let us first consider Eq.\ (\ref{eq:timeindept}) for the case of a localised source, far from the source. Let us introduce spherical coordinates $r$, $\vartheta$, $\varphi$, in which the Laplacian $\nabla^2$ may be written as
\begin{equation}\label{eq:lap}
 \nabla^2 = \frac{\partial^2}{\partial r^2} + \frac{2}{r}\frac{\partial}{\partial r} + \frac{1}{r^2}\nabla^2_\Omega\,,
\end{equation}
where $ \nabla^2_\Omega$
denotes the angular part of the Laplacian. Considering the equation to $O(1/r)$, the angular part may be neglected. In this order, therefore, the angle dependence of field is arbitrary, and one may seek a solution in the form of an exponential times a series in $1/r$. The exponent is determined to be $\pm m_A r$, with the negative sign corresponding to a finite energy solution, and the leading power of $r$ is $-1$,
\begin{equation}\label{eq:asyser}
 A_0 = \frac{\e^{-m_A r}}{r}f(\vartheta,\varphi) + {\cal O}(r^{-2}\e^{-m_A r})\,.
\end{equation}
At this order, the angle dependence is arbitrary; it is, in fact, determined by the source. Most importantly, the scalar potential $A_0$ tends to zero exponentially (it is in fact of order ${\cal O}(\e^{-m_Ar}/r)$) for $r\to \infty$. Let us now consider Gauss' theorem:
\begin{equation}\label{eq:Gauss}
 Q = \int \d^3 r (-m_A^2 A_0 + \varrho_\phi + \varrho_{\rm ext}) = \int \d^3 r \partial_i E_i = \lim_{r\to\infty}r^2 \int_{-\pi}^\pi \d\varphi\int_0^\pi \d\vartheta \sin\vartheta E_r = 0\,,
\end{equation}
where $E_r = -\partial A_0/\partial r$.

The most important consequence of Eq.\ (\ref{eq:Gauss}) is \emph{perfect shielding}: the total induced charge (the density of which is $-m_A^2 A_0 + \varrho_\phi$) exactly cancels the external charge \emph{globally}.

\subsection{Perturbative solution}\label{ssec:lin}
Let us now consider the case when the sources are ``weak'',i.e.\
\begin{equation}\label{eq:weaksrc}
 \rho_{\rm ext} = \epsilon \varrho^{(1)}_{\rm ext}\,,
\end{equation}
one can expand the fields in series of $\epsilon$, as
\begin{equation}\label{eq:expan}
 A_0 = \epsilon A_0^{(1)} + \epsilon^2 A_0^{(2)} + \dots\,,\quad
 \chi = \epsilon^2 \chi^{(2)} + \dots\,,
\end{equation}
[as the external charge is absent from the field equation (\ref{eq:scalarRho}), there is no linear contribution to the scalar field $\chi$].

In each order, the equations assume the form
\begin{equation}\label{eq:HOA0}
 (\nabla^2-m_s^2)\chi^{(k)} = -\xi^{(k)}\,,\quad
 (\nabla^2-m_A^2)A_0^{(k)} = -\sigma_0^{(k)}\,,
\end{equation}
where the source terms $\xi^{(k)}$, $\sigma_0^{(k)}$ are determined by the lower order solutions (up to order $(k-1)$).
\begin{equation}\label{eq:ords}
\begin{aligned}
  \xi^{(1)}\, &= 0\,,\quad &&\sigma_0^{(1)} = \rho_{\rm ext}^{(1)}\,,\\
  \xi^{(2)}\, &= e^2 v A_\mu^{(1)}A^{(1)\mu}\,,\quad &&\sigma_0^{(2)} = -2e^2 v \chi^{(1)} A_0^{(1)}\,,
\end{aligned}
\end{equation}
and the induced charge density, in each order, is given as
\begin{equation}\label{eq:indsrc}
 -m_A^2 A_0^{(k)}+\sigma_0^{(k)}\,.
\end{equation}
It follows from Eqs.\ \eqref{eq:radeq} that $\chi^{(1)}=0$, and therefore $\varrho_\phi^{(1)}=0$ too. In this order, all of the induced charge comes from the term $-m_A^2 A_0^{(1)}$.

The solution is obtained in each order with the help of the Yukawa Green function, $G_i = 1/(4\pi r)\exp(-m_i r)$, $i=s,A$, in the form
\begin{equation}\label{eq:useG}
\begin{aligned}
 A_0^{(k)}(x_i) &= \int \d^3 x' G_A(x_i - x_i') \sigma_0^{(k)}(x_i')\,,
 \quad \quad G_A({\bf x})=\frac{1}{4\pi |{\bf x}|}\exp({-m_A |{\bf x}|})\,,\\
  \chi^{(k)}(x_i) &= \int \d^3 x' G_s(x_i - x_i') \xi^{(k)}(x_i')\,,
 \quad \quad G_s({\bf x}) = \frac{1}{4\pi |{\bf x}|}\exp({-m_s |{\bf x}|})\,.
 \end{aligned}
\end{equation}

The charge in subsequent orders is calculated as
\begin{equation}\label{eq:Q1}
Q^{(1)} = Q^{(1)}_{\rm ext} + Q^{(1)}_{\rm ind}\,,\quad\quad
Q^{(k)}=Q^{(k)}_A + Q^{(k)}_\phi\,,
\end{equation}
where
\begin{equation}
 Q^{(1)}_{\rm ext} = \int \d^3 x \rho_{\rm ext}^{(1)}\,,\quad\quad
 Q^{(k)}_\phi = \int \d^3 x \sigma_0^{(k)}\,,
\end{equation}
and
\begin{equation}\label{eq:QA}
Q^{(k)}_A = -\int \d^3 x m_A^2 A^{(k)} = -m_A^2\int \d^3 x \d^3 x' G_A(x_i-x_i') \sigma_0^{(k)}(x_i') = -Q^{(k)}_\phi\,,
\end{equation}
and similarly $Q^{(1)}_A = -Q^{(1)}_{\rm ext}$,
where the last equality in Eq.\ (\ref{eq:QA}) is obtained by performing the integration over $x$ first.

The procedure for obtaining the full solution in the form of the series in Eq.\ (\ref{eq:expan}) order by order, each order having a source from the lower ones is referred to as a \emph{dressing procedure}. Eq.\ (\ref{eq:QA}) shows, that there is perfect global charge screening, $Q^{(k)}_\phi + Q^{(k)}_A =0$
in each order.

The Yukawa Green function used in Eq.\ (\ref{eq:useG}) also tells us something about \emph{local charge cancellation}. The contribution of a point charge, if linear approximation suffices, is screened within a sphere of radius $O(1/m_A)$. If the external sources do not change significantly on this scale, the external charge \emph{density} is cancelled by the induced charge to a very good accuracy.

Another length scale stems from the scalar, $1/m_s$. In the terminology of superconductivity, the length scale $1/m_A$ is termed \emph{penetration depth} and $1/m_s$ is the \emph{correlation length}. This determines the nature of the interaction between point particles at large distance: if the penetration length is larger, the vector interaction dominates, and the interaction is repulsive, and, on the contrary, if the correlation length is larger, scalar interaction dominates, and the interaction is attractive. The same holds for the interaction between flux tubes in superconductors, and, therefore, this determines the magnetic properties of superconductors; see, e.g., Ref.\ \cite{tinkham}.

\subsection{Spherical symmetry}\label{ssec:sphSer}
Let us also consider spherical symmetry for the series solution, i.e., when $\rho_{\rm exr} = \rho_{\rm ext}(r)$. In this case, the $A_0$ and $\chi$ are also spherically symmetric. Introducing the shorthand notation $y_i^{(k)}$ $i=A,s$ for $A_0^{(k)}$ resp.\ $\chi^{(k)}$ Eqs.\ (\ref{eq:HOA0}) can be compactly written as
\begin{equation}\label{eq:HOAsph}
   \frac{1}{r^2}\left(r^2{y_i^{(k)}}'\right)' -m_i^2 y_i^{(k)} = -h_i^{(k)}\,,
  \end{equation}
where $h_s^{(k)}=\xi^{(k)}$ and $h_A^{(k)}=\sigma_0^{(k)}$.
Eqs.\ \eqref{eq:HOAsph} are second order inhomogeneous linear differential equations, which can be solved using the two linearly independent solutions, $y_{i\pm}=\e^{\pm m_i r}/r$, of the respective homogeneous equations, and their Wronskian $W_i = y_{i+}y_{i-}'-y_{i+}'y_{i-} = -2m_i/r^2$, ($i=A,s$) as
\begin{equation}\label{eq:Wronski}
 y_i^{(k)}(r) = \frac{\e^{-m_i r}}{r}\int\limits^r_{r_0}\!\d x\, \frac{\e^{m_i x}h_i^{(k)}(x)}{W_i(x)}-\frac{\e^{m_i r}}{r}\int\limits^r_\infty\!\d x\, \frac{\e^{-m_i x}h_i^{(k)}(x)}{W_i(x)}\,,\\
\end{equation}
where the integration constants (limits) are chosen to ensure the boundary conditions.
The condition that both $A_0$ and $\chi$ tend to $0$ for $r\to\infty$ is clearly implemented in Eqs.\ \eqref{eq:Wronski}.

The boundary conditions at $r=0$ come from regularity in the sense that terms $\propto 1/r$ at $r\to 0$ be absent (otherwise they would yield unwanted Dirac-delta sources in $\nabla^2 A_0$ or $\nabla^2\chi$), this fixes the value of the integration constant ($r_0$).

\section{Point source}\label{sec:ptsrc}
A point source is defined by a Dirac delta as the external charge density,
\begin{equation}\label{eq:PTsrc}
\varrho_{\rm ext}({\bf x}) = q\delta^{(3)}({\bf {\bf x}}) \,.
\end{equation}
In this case, we shall use $q$ as the expansion parameter, $\epsilon=q$.

Assuming spherical symmetry the leading term is
\begin{equation}\label{eq:ptA0}
 A_0^{(1)}(r) = G_A(r)=\frac{\e^{-m_{\scriptscriptstyle{A}}r}}{4\pi r}\,.
\end{equation}
The energy of the point particle solutions obviously diverges since for $r\to 0$, $G\sim 1/(4\pi r)$, the electromagnetic field contribution to the energy density, ${\bf E}^2/2 \sim 1/(32\pi^2 r^4)$, thus $E = 4\pi\int \mathcal{E}r^2\d r$, which is divergent.

The second order contribution to the scalar field, $\chi^{(2)}(r)$ is obtained from Eq.\ \eqref{eq:Wronski}
yielding
\begin{equation}\label{eq:PT2}
\begin{aligned}
 \chi^{(2)}(r) &= -\frac{e^2 v}{2(4\pi)^2 m_s r}\Big[ \e^{-m_s r}\left(\mathop{\rm Ei}[(m_s-2m_A)r]-\log\frac{|m_s-2m_A|}{m_s+2m_A}\right)\\
 &\quad\quad\quad\quad\quad\quad\quad- \e^{m_s r}\mathop{\rm Ei}[-(m_s+2m_A)r] \Big]\,,\\
 \mathop{\rm Ei}(x) &= -\hspace{-1.2em}\int\limits_{-\infty}^x\! \d t\,\frac{\e^t}{t}\,.
\end{aligned}
\end{equation}
where $-\hspace{-.9em}\int$ denotes the principal value integral, see Refs.\cite{AS, NIST}. In deriving $\chi^{(2)}(r)$
we have also used the expansion $\mathop{\rm Ei}(x) \sim -\gamma + \ln (\vert x\vert) +x$ for $x\to 0$.
From \eqref{eq:PT2} it follows that for $r\to 0$  $\chi^{(2)}(r)\propto \log(r)$.

In the special case $m_s= 2m_A$, the solution in \eqref{eq:PT2} is replaced by
\begin{equation}\label{eq:PT21}
\chi^{(2)}(r)=-\frac{e^2 v}{2(4\pi)^2 m_s r}\Big[\e^{-m_sr}\ln(r/r_0)-\e^{m_sr}\mathop{\rm Ei}[-2m_sr]
\Big]\,,\quad r_0=\frac{\e^\gamma}{2m_s}\,.
\end{equation}

\subsection{Interaction energy}\label{ssec:inten}

The interaction energy between two point charges $q_1$, $q_2$ placed at $\vec{r}_1$ resp.\ $\vec{r}_2$ at a separation $\vec{r}=\vec{r}_1-\vec{r}_2$ in both (massive) electrodynamics and in Klein-Gordon theory is calculated as $\pm 1$ times the product of the field of one charge at the position of the second one multiplied by the value second charge (see Refs.\ \cite{LL2, IS}and Appendices \ref{app:KGInt} and \ref{app:PInt}).

In the case $m_s > m_A$, it is the field due to $A_0^{(1)}$ which dominates for large $r$, yielding
\begin{equation}\label{eq:ForceTypeI}
 V_{\rm II}(r) = \frac{q_1 q_2}{4\pi r}\e^{-m_A r}\,,
\end{equation}
whereas for $m_A>m_s$, it is the scalar field $\chi^{(2)}$ which dominates. In this case, for $r\to \infty$, the leading contribution is given by
\begin{equation}\label{eq:PT2asy}
 \chi^{(2)}(r) \sim \frac{e^2 v}{2(4\pi)^2 m_s r}\e^{-m_s r}\log\frac{|m_s-2m_A|}{m_s+2m_A}\,.
\end{equation}
The field $\chi^{(2)}$ in Eq.\ (\ref{eq:PT2asy}) is the field of a point source in a Klein-Gordon field with strength $e^2 v/2/(4\pi)/m_s \log|m_s-2m_A|/(m_s+2m_A)$.

To obtain the interaction potential between $q_1$ and $q_2$, the field of charge $q_2$ is approximated as emanating from a point source determined by the near field limit (this approximation has been used by Ref.\ \cite{speight} for the interaction energy of vortices in superconductors). That is $\xi^{(2)}$ is replaced by $q_s^{(2)}\delta^3(\vec{r}_2)$, where $q_s^{(2)}=-\int \d^3 x \xi^{(2)} = -e^2 v/(8\pi m_A)$. $q_s^{(2)}$ multiplies the field of the charge $q_1$, yielding
\begin{equation}\label{eq:ForceTypeII}
V_{\rm I}(r) = \frac{e^4 v^2 q_1^2 q_2^2}{4(4\pi)^3 m_s m_A}\log\frac{2m_A-m_s}{2m_A+m_s}\frac{\e^{-m_s r}}{r}\,.
\end{equation}
The indices ``I'' and ``II'' to distinguish $m_A>m_s$ and $m_s>m_A$ ($\beta <$ or $>\sqrt{2}$) were choosen in accord with the terminology of superconductivity \cite{tinkham}. In both cases, the force is exponentially decreasing with the separation; however, in the case of a type I setting, \emph{like charges attract} (the logarithm is negative). Formulae (\ref{eq:ForceTypeI}) and (\ref{eq:ForceTypeII}) are valid for large separations $r$.

\subsection{Numerical calculations}\label{ssec:ptnum}

The Dirac delta source does not appear in the radial equations, 
which are only defined for $r>0$. Instead, it manifests itself in the boundary conditions \cite{adler, adlerpiran, ishiharaogawascreen}, as $r\to 0$, $A_0 \sim q/(4\pi r)$, i.e., very close to the source, it is unshielded. For a series solution, the leading power for a scalar field is determined from considering the coefficient of the lowest power, yielding
\begin{equation}\label{eq:bdryPtr0}
 r\to 0\,: \quad A_0 \sim \frac{q}{4\pi r}\,,\quad \chi\sim \chi_0 r^\gamma\,,
\end{equation}
with $\gamma=-1/2 + \sqrt{1/4-\kappa^2}$ (for $\kappa<1/2$), where $\kappa=eq/(4\pi)$ and
\begin{equation}\label{eq:bdryPtr0LQ}
 r\to 0\,: \quad A_0 \sim \frac{q}{4\pi r}\,,\quad \chi\sim \chi_0 \frac{1}{\sqrt{r}}\cos(\gamma_1 \log r+\delta)\,,
\end{equation}
where $\gamma_1 = \sqrt{\kappa^2-1/4}$ for $\kappa>1$. In eqs.\ (\ref{eq:bdryPtr0}) and (\ref{eq:bdryPtr0LQ}), $\chi_0$ is a constant determined from the numerical solution of the radial equations. 

For $r\to \infty$, all fields approach their vacuum value,
\begin{equation}\label{eq;bdryrInf}
 r\to\infty\,:\quad A_0 \to 0\,,\quad \chi\to 0\,.
\end{equation}

\begin{figure}[h!]
\begin{subfigure}[t]{0.5\textwidth}
 \noindent\hfil\includegraphics[scale=.5]{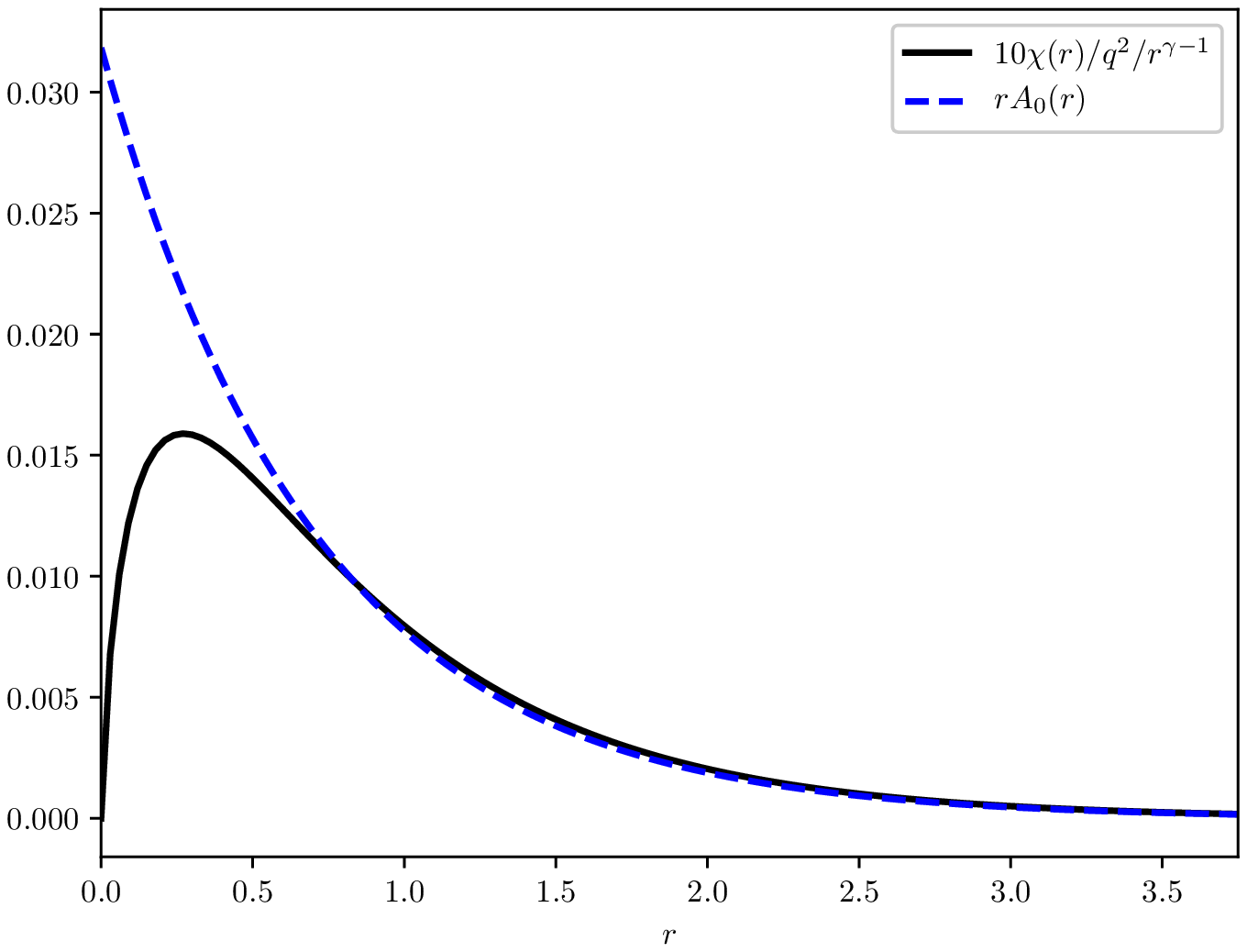}
 \caption{}
 \label{fig:ptsrcf}
\end{subfigure}
\begin{subfigure}[t]{0.5\textwidth}
 \noindent\hfil\includegraphics[scale=.5]{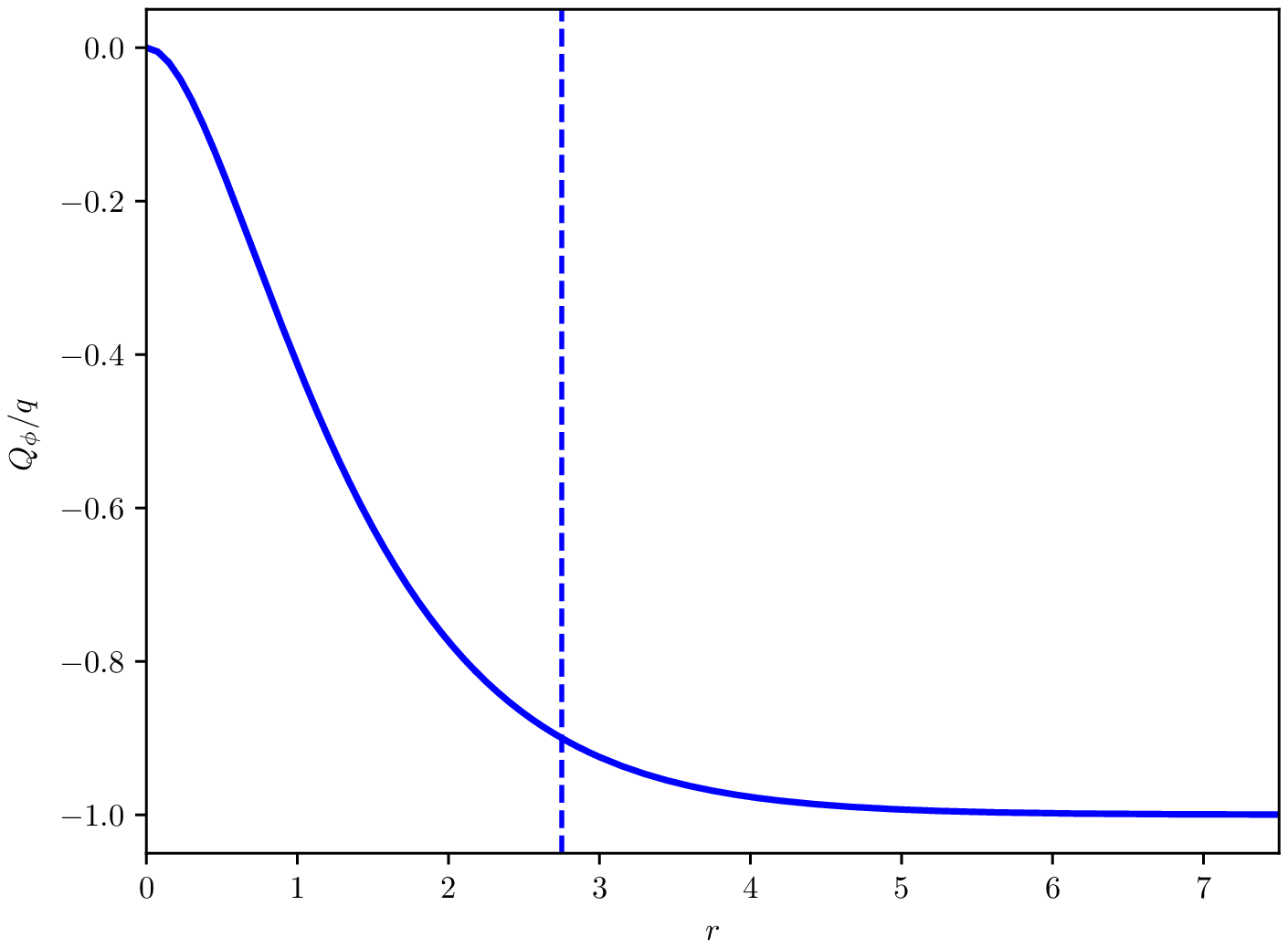}
 \caption{}
\label{fig:ptsrcQ}
\end{subfigure}
\caption{(a) The profile functions $A_0$ and $\chi$ of a solution for point charge; $\beta=2.0$, $q=0.4$. (b) The charge distributions of the same solution. The dashed vertical line shows $r=R_c$.}
\label{fig:ptsrc}
\end{figure}

We have calculated numerical solutions using the {\sc Colnew} package \cite{colnew, ascher}. An example numerical solution is diplayed in Fig.\ \ref{fig:ptsrcf}, and the corresponding charge in Fig.\ \ref{fig:ptsrcQ}. For the given value of the charge $q=0.4$ and self-interaction $\beta=2.0$, the leading order series agrees extremely well with the exact result (within line width of Fig.\ \ref{fig:ptsrcf}).

\begin{figure}[h!]
\begin{subfigure}[t]{0.5\textwidth}
 \noindent\hfil\includegraphics[scale=.5]{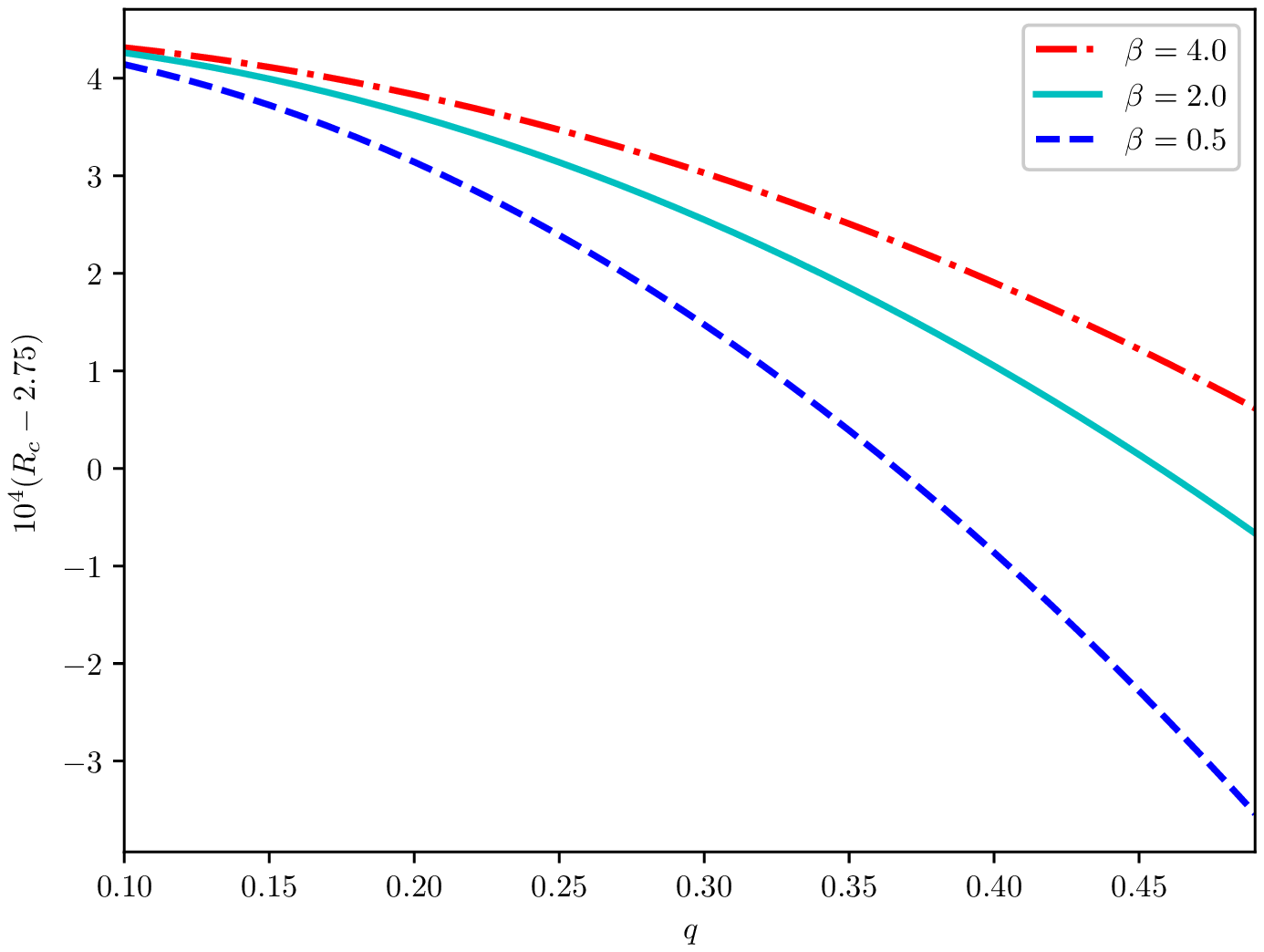}
 \caption{}
 \label{fig:ptRq}
\end{subfigure}
\begin{subfigure}[t]{0.5\textwidth}
 \noindent\hfil\includegraphics[scale=.5]{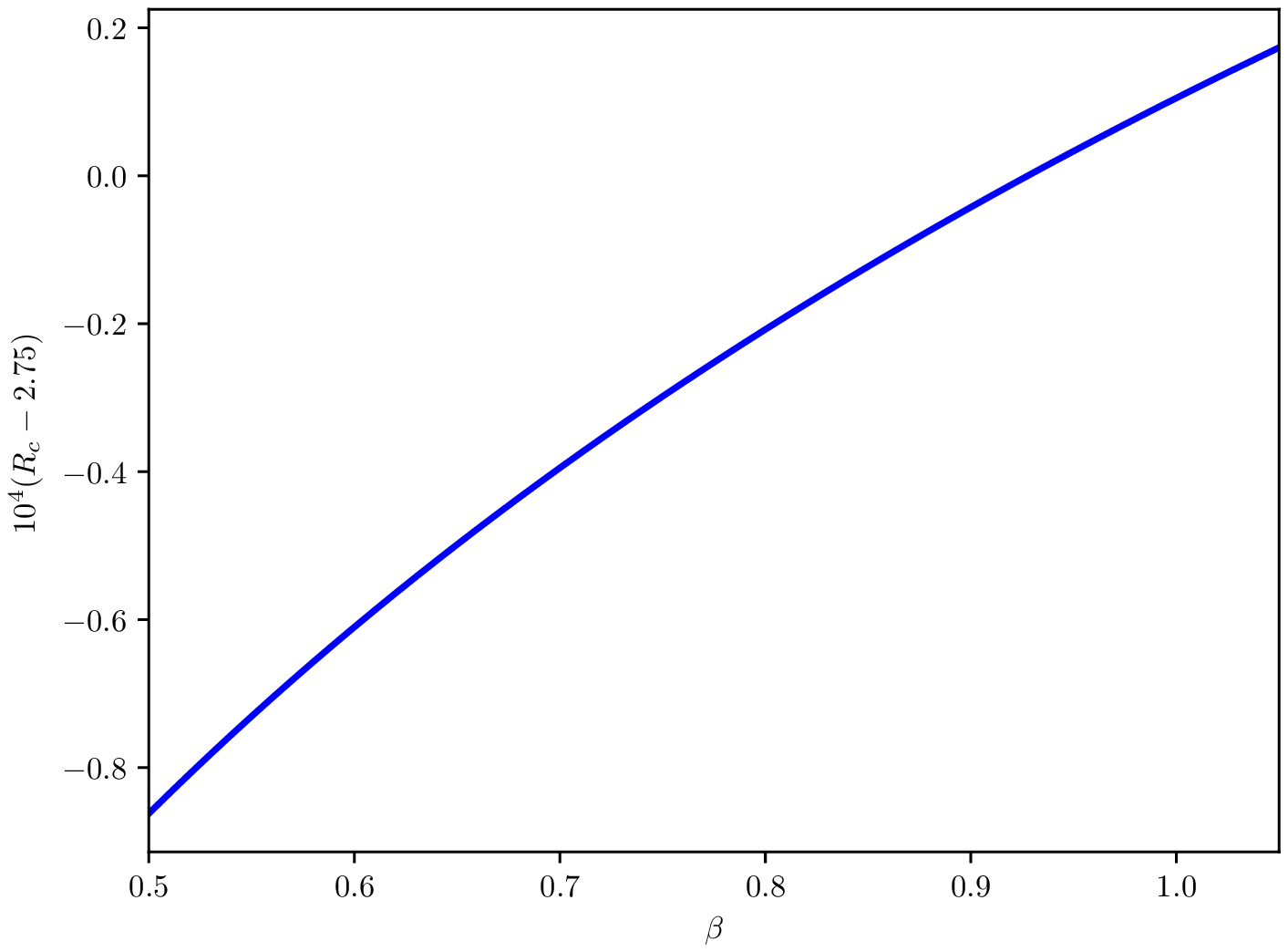}
 \caption{}
\label{fig:ptRb}
\end{subfigure}
\caption{Point source: the dependence of the radius of screening with a precision of 0.9 on (a) the charge $q$ and (b) the strength of self-interaction $\beta$ (for $q=0.4$).}
\label{fig:ptR}
\end{figure}

In Fig.\ \ref{fig:ptR} we have plotted the radius $R_c$, defined as
\begin{equation}\label{eq:Rc}
 Q_\phi(R_c) = -0.9 q\,,
\end{equation}
as a function of $q$ and $\beta$. We have found, that $R_c$ is a decreasing function of $q$ and an increasing one of $\beta$.

\section{Gaussian source}\label{sec:Gauss}
Another interesting example is a Gaussian source, with source radius $R_s$,
\begin{equation}\label{eq:GaussSrc}
 \rho_{\rm ext} = \frac{q}{(2\pi)^{3/2}R_s^3}\e^{-\frac{r^2}{2R_s^2}}\,.
\end{equation}
Again, we shall use $\epsilon=q$ as the expansion parameter.
The integration constant in the solution given in Eq.\ (\ref{eq:Wronski}) is chosen to ensure regularity for $r\to 0$ (i.e.\ to cancel the $1/r$ terms), yielding
\begin{equation}\label{eq:LinGauss}
 A_0^{(1)} = \frac{1}{4\pi r}\e^{-m_A r}\frac{\e^{\frac{m_A^2 R_s^2}{2}}}{2}\left[ \mathop{\rm Erfc}\left(\frac{m_A R_s^2-r}{\sqrt{2}R_s}\right)- \e^{2 m_A r}\mathop{\rm Erfc}\left(\frac{m_A R_s^2+r}{\sqrt{2}R_s}\right)\right]\,,
\end{equation}
where $\mathop{\rm Erfc}(x)=1-\int_0^x \d x' \exp(-x^2)=1-\mathop{\rm Erf}(x)$ is the complementary error function (see Ref.\ \cite{RG}).

The expression for the potential $A_0^{(1)}$ of the Gaussian source in Eq.\ (\ref{eq:LinGauss}) starts with the potential $G$ of the point source, see Eq.\ (\ref{eq:Yukawa}). It is instructive to introduce an effective charge as
\begin{equation}\label{eq:EffCharge}
 Q_{\rm eff} = q\lim_{r\to\infty}\frac{A_0^{(1)}}{G} = q\e^{\frac{m_A^2 R_s^2}{2}} \approx q \left(1+ \frac{m_A^2 R_s^2}{2}+\dots\right)\,,
\end{equation}
where we have used the fact, that $\mathop{\rm Erf}(x)\sim 1 + \exp(-x^2)/(\sqrt{\pi}x)$ for $x\to\infty$ (7.1.23 and 7.12.1 in Refs.\ \cite{AS, NIST}, resp.), i.e., the leading contribution comes from the constant $2$ term in the asymptotics of the first $\mathop{\rm Erfc}$ function in Eq.\ (\ref{eq:LinGauss}).

According to Eq.\ (\ref{eq:EffCharge}), at a large distance, the potential due to an extended source is stronger than that of a point charge of the same magnitude. This is physically plausible, as for massless electrodynamics, they agree, and here, some part of the source is less strongly screened.

\begin{figure}[h!]
\begin{subfigure}[t]{0.5\textwidth}
 \noindent\hfil\includegraphics[scale=.5]{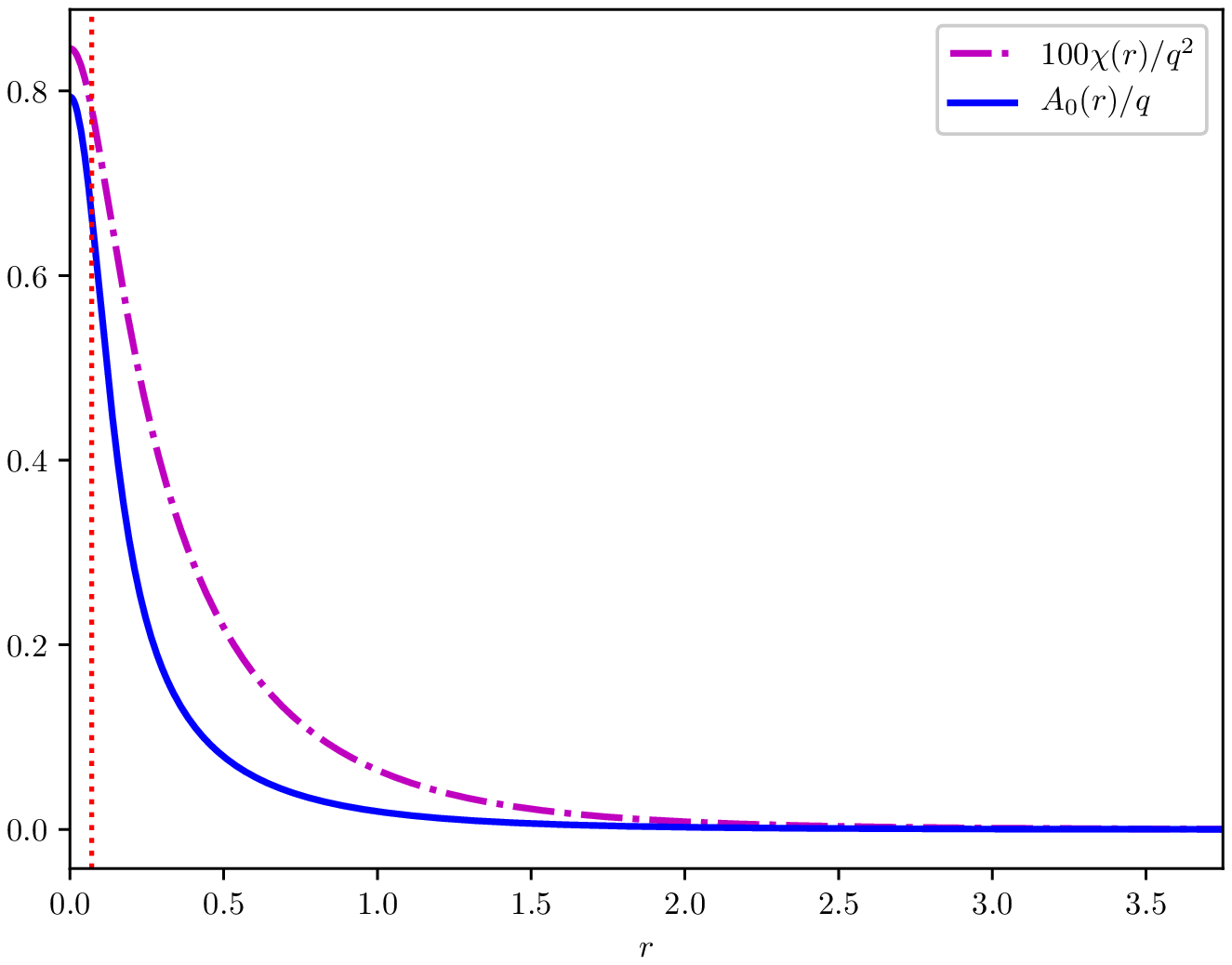}
 \caption{}
 \label{fig:xtsrcf}
\end{subfigure}
\begin{subfigure}[t]{0.5\textwidth}
 \noindent\hfil\includegraphics[scale=.5]{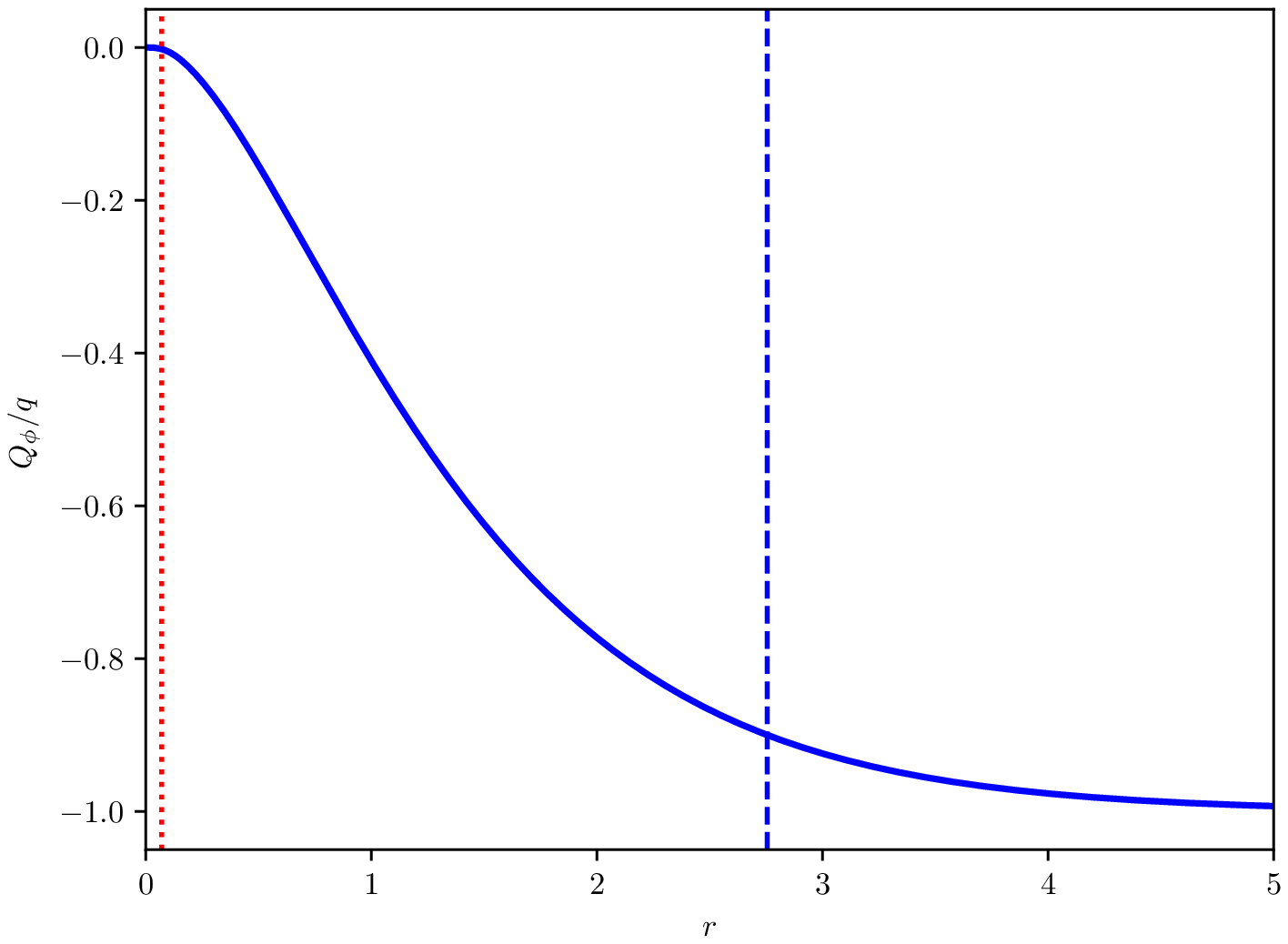}
 \caption{}
\label{fig:xtsrcQ}
\end{subfigure}
\caption{(a) The profile functions $A_0$ and $\chi$ of a solution for the Gaussian charge distribution; $\beta=2.0$, $q=0.4$, $R_s=0.1/\sqrt{2}$. (b) The charge distributions of the same solution. The dotted vertical line shows $r=R_s$ and the dashed one $r=R_c$.}
\label{fig:xtsrc}
\end{figure}

\begin{figure}[h!]
\begin{subfigure}[t]{0.5\textwidth}
 \noindent\hfil\includegraphics[scale=.5]{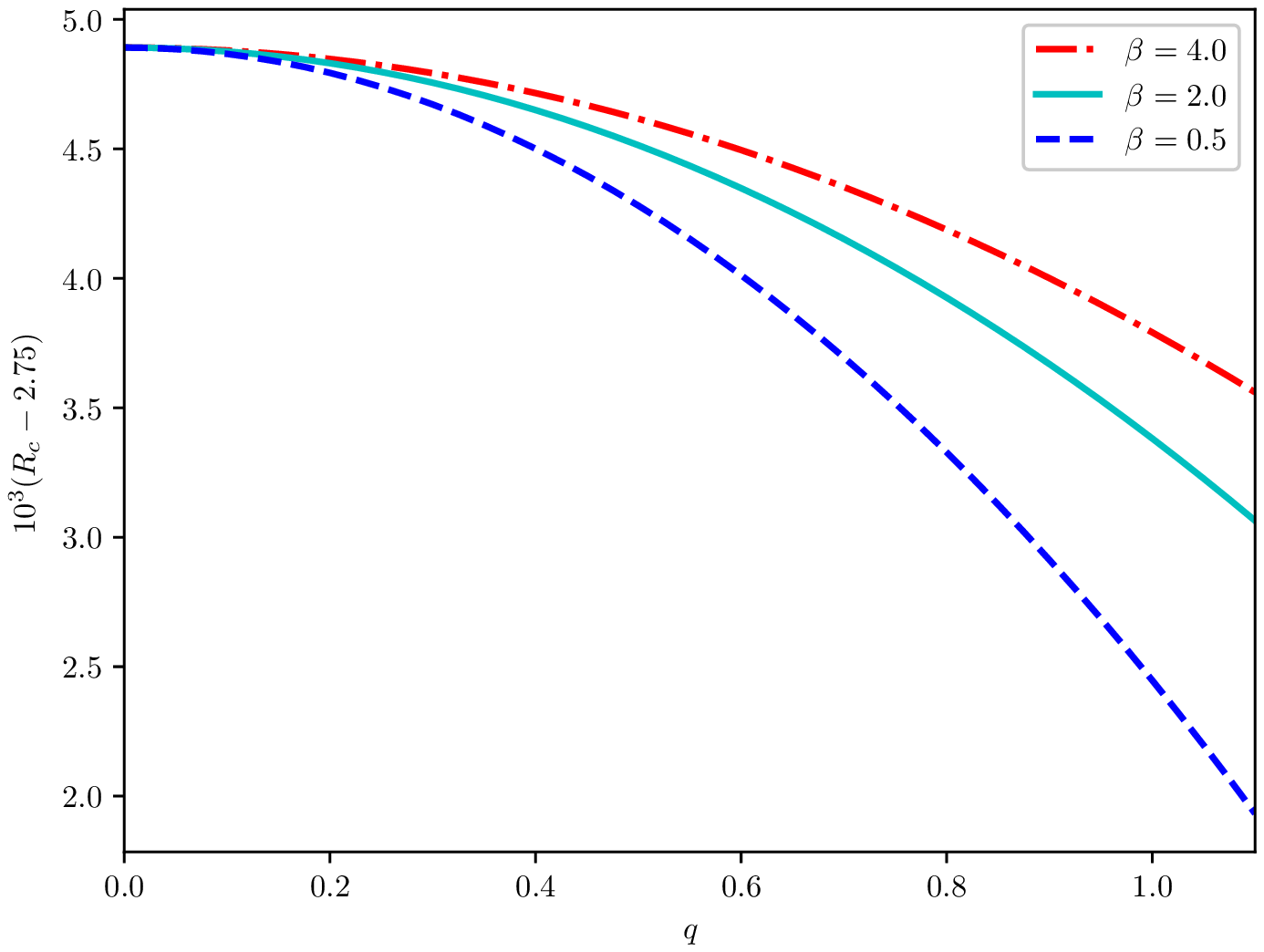}
 \caption{}
 \label{fig:xtRq}
\end{subfigure}
\begin{subfigure}[t]{0.5\textwidth}
 \noindent\hfil\includegraphics[scale=.5]{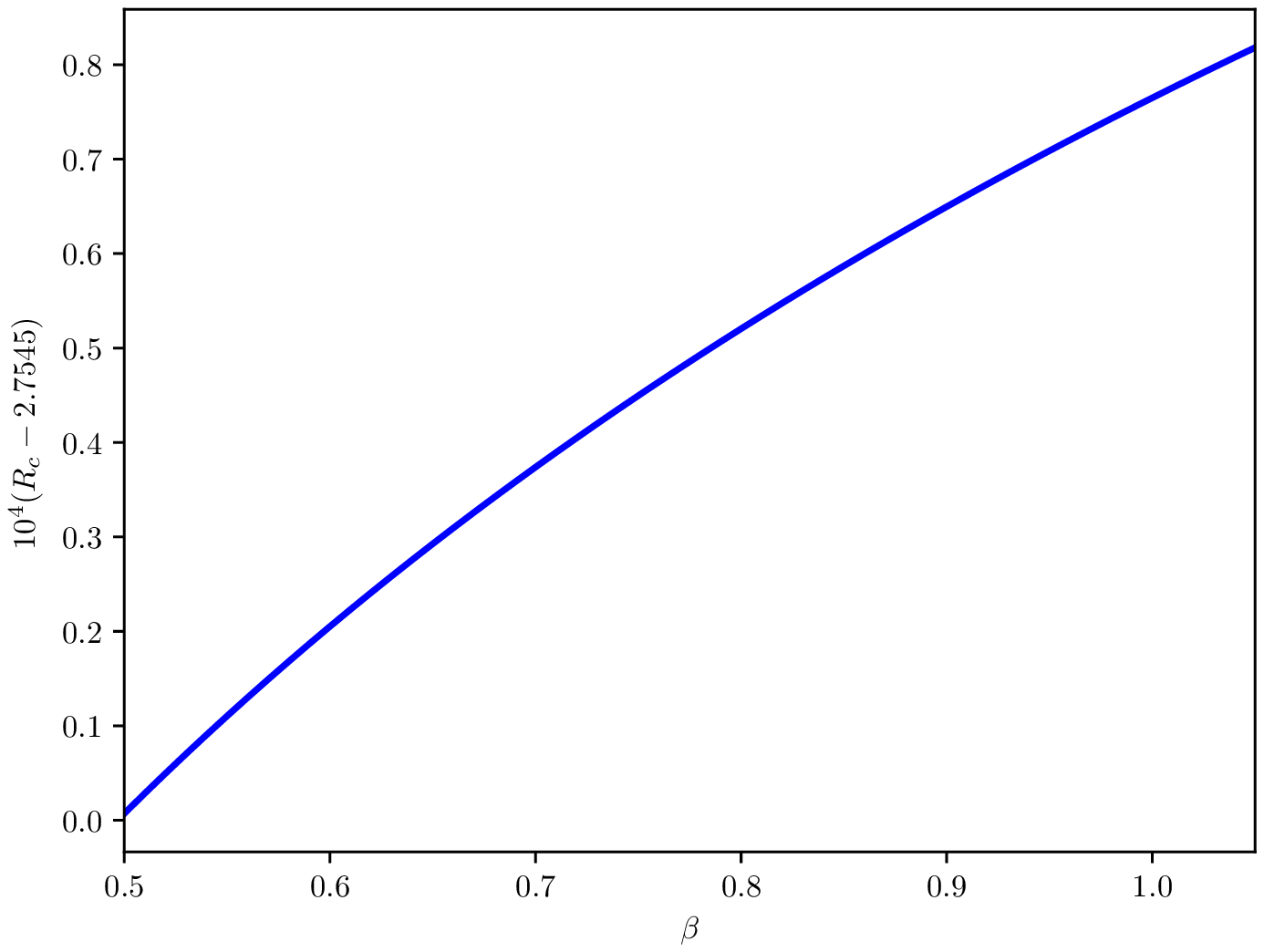}
 \caption{}
\label{fig:xtRb}
\end{subfigure}
\caption{Gaussian source: the dependence of the radius of screening with a precision of 0.9 on (a) the charge $q$ and (b) the strength of self-interaction $\beta$ (for $q=0.4$). In both cases, $R_s=1/10/\sqrt{2}$.}
\label{fig:xtR}
\end{figure}

We have also calculated numerical solutions for the extended source distribution given by Eq.\ (\ref{eq:GaussSrc}), using the {\sc Colnew} package \cite{colnew, ascher}. A typical solution is shown in Fig.\ \ref{fig:xtsrc}. On the figures, the half-width of the source $R_s$ and the radius $R_c$ where the screening reaches an accuracy of $0.9$ are also shown.
Again an excellent agreement is found between the first order analytical solution (\ref{eq:LinGauss}) and the numerical one depicted in Fig.\ \ref{fig:xtsrcf}, the difference between the two curves being smaller than the line width. We have also used the numerical solutions to obtain $R_c$ as a function of $q$, and have found that it is monotonically decreasing (Fig.\ \ref{fig:xtRq}) and of $\beta$ (Fig.\ \ref{fig:xtRb}), of which it is weakly increasing.
Our results are in perfect agreement with the numerical results of Ref.\ \cite{ishiharaogawascreen}. In the case of closed form formulae, our analyses complement each other. Ref.\ \cite{ishiharaogawascreen} consider the case of a small source, $R_s\ll 1/m_A$, and in this case present an expansion of $A_0$ in $r$, and a large source, $R_s\gg 1/m_A$ an in this case show that charge cancellation is local, and $A_0 \approx \rho_{\rm ext}/m_A^2$ in the leading order. These approximations are found to be of good agreement with numerical results. Our analysis, on the other hand, provides the expansion of the solution in the external charge $q$, and is also found to be in excellent agreement with numerical results.

\section{Homogeneous sphere}\label{sec:hom}
\begin{figure}[h!]
\begin{subfigure}[t]{0.5\textwidth}
 \noindent\hfil\includegraphics[scale=.5]{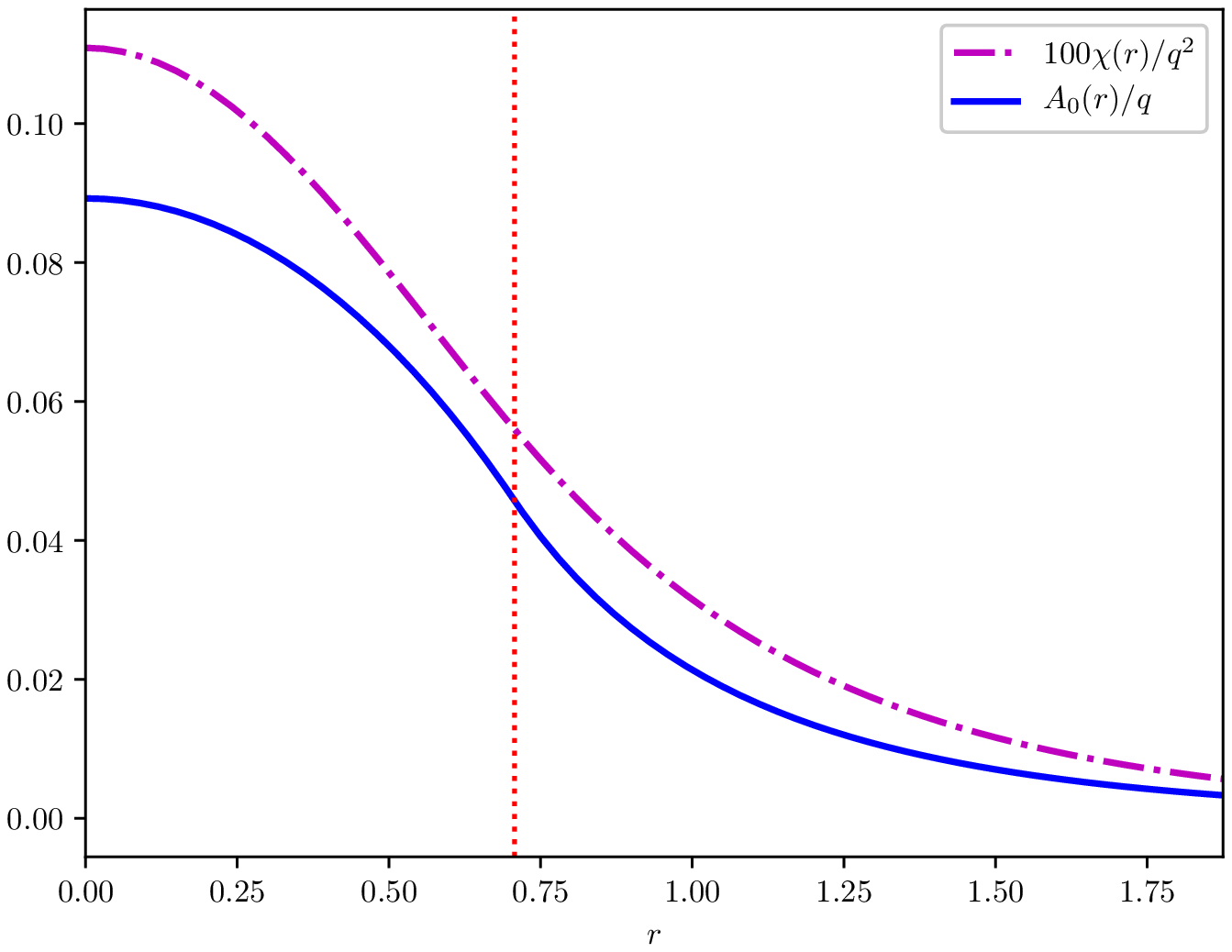}
 \caption{}
 \label{fig:hsrcf}
\end{subfigure}
\begin{subfigure}[t]{0.5\textwidth}
 \noindent\hfil\includegraphics[scale=.5]{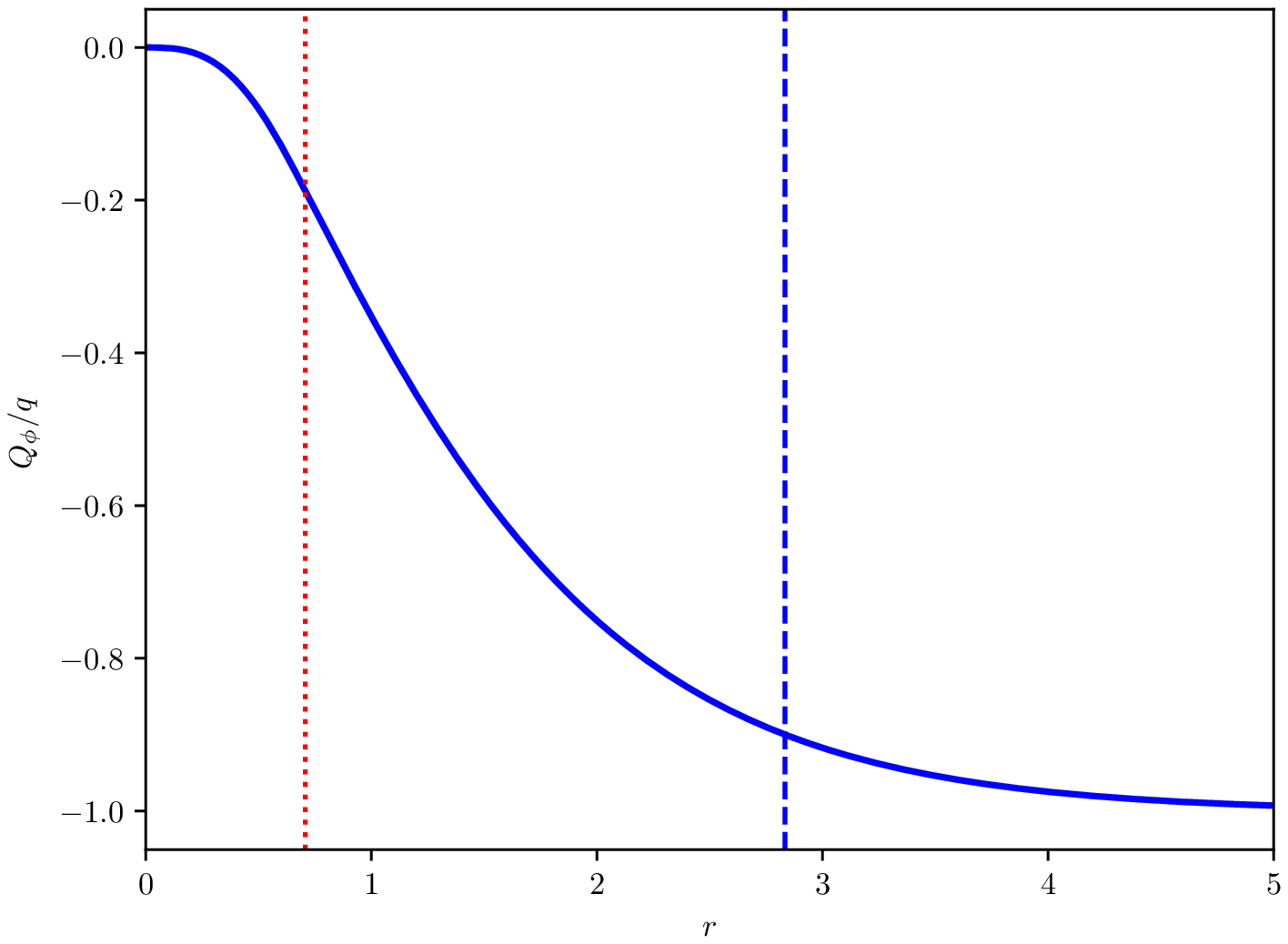}
 \caption{}
\label{fig:hsrcQ}
\end{subfigure}
\caption{(a) The profile functions $A_0$ and $\chi$ of a solution for the homogeneous charge distribution; $\beta=2.0$, $q=0.4$, $R_s=1/\sqrt{2}$. (b) The charge distributions of the same solution. The dotted red line shows $r=R_s$ and the dashed blue one $r=R_c$.}
\label{fig:hsrc}
\end{figure}

Let us also consider a homogeneous sphere as the source, $\rho_{\rm ext} = \rho_0 \Theta(R_s-r)$, where $\Theta(x)$ is the Heaviside function, $\Theta(x<0)=0$, $\Theta(x>0)=1$. In this case, obtaining the scalar potential using formula (\ref{eq:Wronski}) yields
\begin{equation}\label{eq:hsrcA0}
\begin{aligned}
 A_0^{(1)}(r) &= \frac{\rho_0/q}{2 m_A^3 r}\big[ \e^{-m_A r}\left( -\e^{m_A r_<}(1-m_A r_<) +\e^{-m_A R_s}(1+m_A R_s)\right)\\
 &\quad\quad\quad\quad + \e^{m_A r}\Theta(R_s-r)\left( \e^{-m_A r}(1+m_A r)-\e^{-m_A R_s}(1+m_A R_s)\right)\big]\,,
\end{aligned}
\end{equation}
where $r_< = \mathop{\rm min}(r,R_s)$. Again, we have used the total charge, $q=4\pi R_s^3 \rho_0/3$ as the expansion parameter.  The agreement with the numerical result is again near perfect, within the line width of Fig.\ \ref{fig:hsrcf}.

The perturbative result,  Eq.\ (\ref{eq:hsrcA0}) allows one to introduce an effective charge again, by comparing the asymptotic form of the solution to $Q_{\rm eff}/(4\pi r)\e^{-m_A r}$, which yields
\begin{equation}\label{eq:hsrcEffCharge}
 Q_{\rm h,eff} = \frac{2\pi\rho_0}{m_A^3}\left( \e^{-m_A R_s}(1+m_A R_s)-\e^{m_A R_s}(1-m_A R_s)\right)\approx q\left(1+\frac{m_A^2 R_s^2}{10}+\dots\right)\,,
\end{equation}
again, the effective charge is slightly larger than the total external charge.

\begin{figure}[h!]
\begin{subfigure}[t]{0.5\textwidth}
 \noindent\hfil\includegraphics[scale=.5]{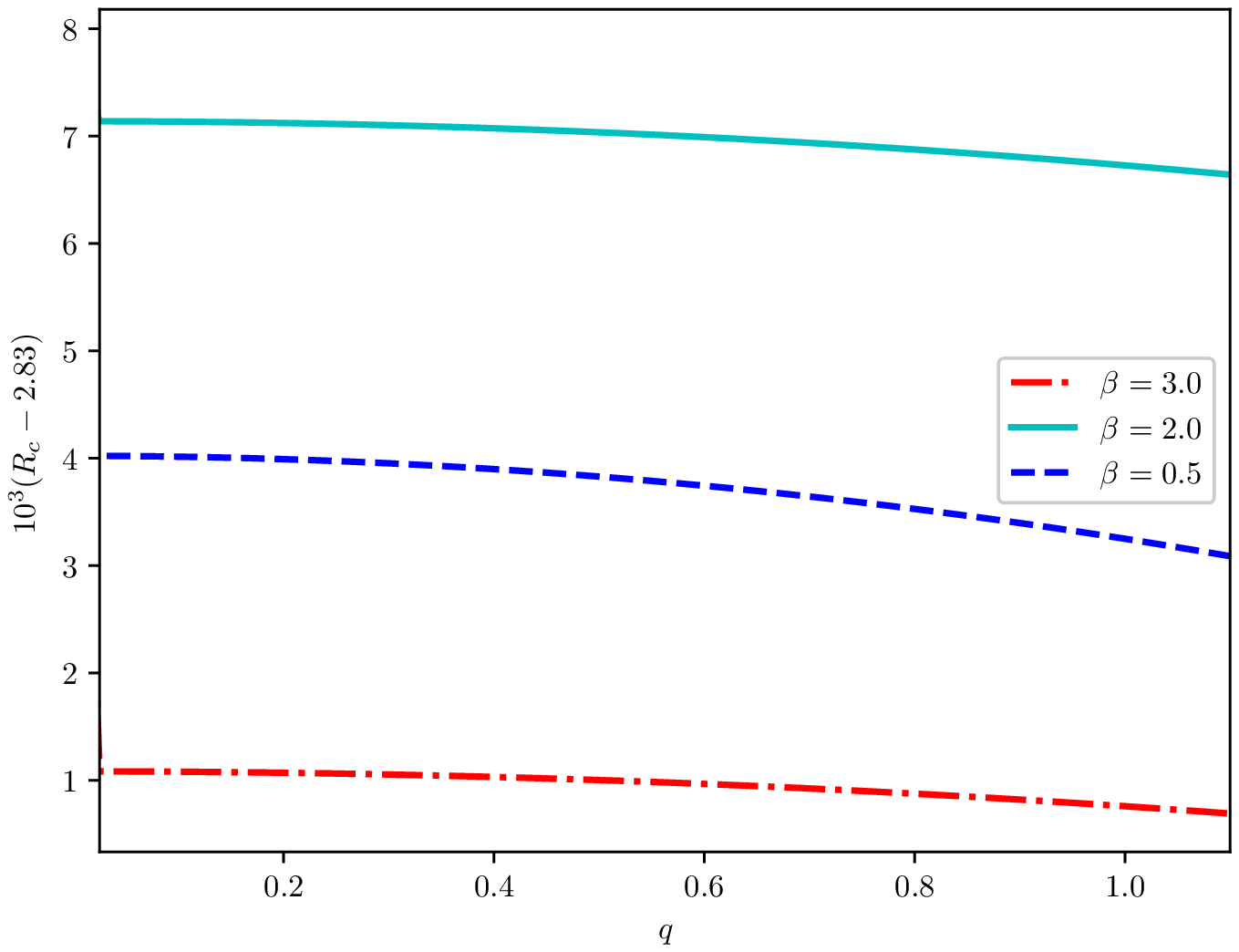}
 \caption{}
 \label{fig:homRq}
\end{subfigure}
\begin{subfigure}[t]{0.5\textwidth}
 \noindent\hfil\includegraphics[scale=.5]{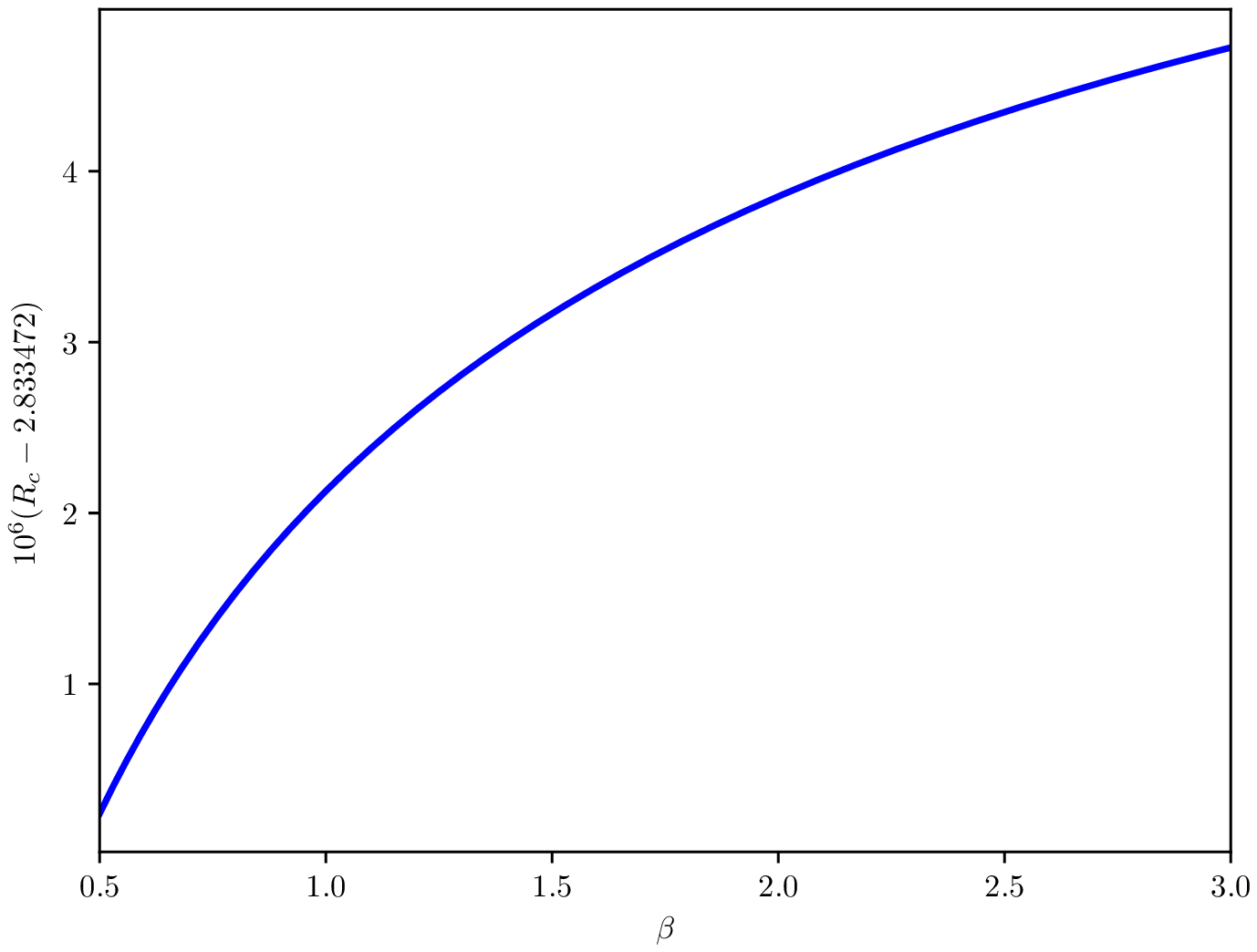}
 \caption{}
\label{fig:homRb}
\end{subfigure}
\caption{Homogeneous sphere source: the dependence of the radius of screening with a precision of 0.9 on (a) the charge $q$ and (b) the strength of self-interaction $\beta$ (for $q=0.1$). In both cases, $R_s=1/\sqrt{2}$.}
\label{fig:homR}
\end{figure}

The behavior of the screening radius $R_c$ on the parameters is similar as before: it is weakly decreasing with $q$ and increasing with $\beta$. (See Fig.\ \ref{fig:homR}.)

In Ref.\ \cite{ishiharaogawascreen}, Eq.\ (\ref{eq:hsrcA0}) has been derived with a matching procedure. In addition, they consider the case of a lage sphere, in which case the fields inside the sphere are constant, and can be found by neglecting derivatives in the radial equations (\ref{eq:radeq}). The latter approximation is also used in Ref.\ \cite{ishiharaogawascreen} for a thick walled spherical source. Ref.\ \cite{ishiharaogawascreen} finds good agreement between the approximate formulae and numerical results.

\section{Conclusions}
The screening of time-independent external charges in the Abelian Higgs model has been considered, from multiple aspects. On one hand, using the asymptotic form of the gauge field far from the charges, it has been shown, from Gauss' law, that the charge screening is exact, the charges induced in the fields of the Abelian Higgs model globally cancel the external charge exactly. The same result has been demonstrated using perturbation theory considering the charge as the perturbation parameter. Local cancellation occurs if the external charge distribution does not change significantly on the scale of $1/m_A$, where $m_A$ is the gauge boson mass. We have obtained simple explicit formulae for the fields in the first nontrivial order in perturbation theory and compared them to numerical results, and found very good agreement.
We note that our results are in perfect agreement and complementing those of Ref.\ \cite{ishiharaogawascreen}.

\paragraph{Acknowledgements} Á.L.\ acknowledges the support of  the Spanish Ministerio de Ciencia, Innovación y Universidades (Grant No.\ PCI2018-092896) and the EU (QuantERA CEBBEC).

\paragraph{Note added in proof} After the completion of the present paper, it has been drawn to our attention, that screening with a mechanism similar to the one in the Abelian Higgs model has
also been considered in Chern-Simons-Higgs theories in Refs.\ \cite{Lee1} and in Bose liquids in Ref.\ \cite{Lee2}.

\appendix
\section{The Yukawa Green function}\label{app:Yukawa}
The linearised equation is
\begin{equation}\label{eq:linA0}
 (\nabla^2-m^2)\psi = -\sigma\,,
\end{equation}
with $\psi$ being $A_0^{(n)}$ or $\chi^{(n)}$ and $\sigma$ its respective source, and $m=m_A$ or $m_s$.

The solution of Eq.\ (\ref{eq:linA0}) may be constructed from the solution of the equation
\begin{equation}\label{eq:linG}
 (\nabla^2-m^2)G = -\delta^{(3)}(x_i)\,,
\end{equation}
in the form
\begin{equation}\label{eq:useGA}
 \psi(x_i) = \int \d^3 x' G(x_i - x_i')\sigma(x_i')\,.
\end{equation}
The function $G$ is the Green function of the operator $\nabla^2-m^2$. It may be computed by seeking it in the form of a Fourier integral,
\begin{equation}\label{eq:Four}
 G({\bf x}) = \int \frac{\d^3 k}{(2\pi)^3} \tilde{G}({\bf k}) \e^{\imagi {\bf k x}}\,.
\end{equation}
Using the fact, that $\delta(x) = \int \d k/(2\pi)\exp(\imagi k x)$, we learn that
\begin{equation}\label{eq:Gt}
 \tilde{G}(k) = -\frac{1}{{\bf k}^2+m^2}\,.
\end{equation}
We may choose coordinates in $k$-space in such a way, that its $k_z$ axis is along the direction of the vector ${\bf x}$, and use spherical coordinates, resulting in
\begin{equation}\label{eq:Four2}
 G = \int \frac{\d k \d\vartheta\d\varphi k^2 \sin\vartheta}{(2\pi)^3}\frac{-1}{k^2+m^2}\e^{\imagi k r \cos\vartheta}\,,
\end{equation}
where $r=|{\bf x}|$. The integral over $\varphi$ merely cancels one of the $2\pi$ factors. The integral over $\vartheta$ can be performed by noting that
$\d\cos\vartheta=-\sin\vartheta\d\vartheta$, and flipping the limits, yielding
\begin{equation}\label{eq:Four3}
 G = \imagi\int_0^\infty \frac{\d k k}{2\pi^2 r}\sin(kr) = \imagi\int_0^\infty \frac{\d k k}{4\pi^2 r}\frac{-1}{k^2+m^2}\e^{\imagi kr}\,.
\end{equation}
Considering now complex values of $k$, and closing the integration contour with a large semicircle on the upper half-plane, its contribution vanishes exponentially, because $r>0$. The direction is positive (counter-clockwise), therefore one needs to add up the residues of poles on the upper half-plane, with a prefactor of $2\pi i$. There is one such pole, at $k=\imagi m$, yielding
\begin{equation}\label{eq:Yukawa}
 G = \frac{1}{4\pi r}\e^{-m r}\,,
\end{equation}
which is nothing else than the Yukawa potential.

\section{The retarded Green function of the massive Klein-Gordon equation}\label{app:KGGreen}
In this appendix, the Green function of the massive Klein-Gordon equation, satisfying
\begin{equation}\label{eq:mKGG}
 (\partial_\mu\partial^\mu + m^2)G_4 = \delta(t)\delta^{(3)}({\bf x})\,,
\end{equation}
is calculated, again, with the help of a Fourier transformations, based on the discussion in Ref.\ \cite{IS},
\begin{equation}\label{eq:Four4}
 G_4 ({\bf x},t) = \int\frac{\d k_0 \d^3 k}{(2\pi)^4}\e^{-\imagi(k_0 t-{\bf k x})}\tilde{G}_4(k_0,{\bf k})\,.
\end{equation}
Using the Fourier representation of the Dirac delta, $\tilde{G}_4=-1/(k_0^2-{\bf k}^2 -m^2)$ is obtained, and, using spherical coordinates for ${\bf k}$, with the 3rd axis aligned along ${\bf r}$, the form
\begin{equation}\label{eq:Four4a}
 G_4 = \int \frac{\d k_0 \d k k^2 \d(\cos\vartheta)\d\varphi}{(2\pi)^4}\frac{-1}{k_0^2-k^2-m^2}\e^{-\imagi k_0 t}\e^{\imagi k r \cos\vartheta}
\end{equation}
is obtained. The integral over $k_0$ is performed using the theorem of residues. For $t>0$, the contour may be closed with a large semicircle on the lower, whereas for $t<0$, in the upper half-plane. Causality demands, that the poles at $k_0 = \pm \sqrt{k^2+m^2}$ be shifted slightly downwards, so that $G({\bf x},t) = 0$ for $t<0$ results. For $t>0$, the integration along the contour is clockwise, pole contibutions are multiplied by $2\pi/i$;
\begin{equation}\label{eq:Four4b}
 G_4 = \Theta(t)\int \frac{\d k k^2 \d(\cos\vartheta)\d\varphi}{(2\pi)^3}\e^{\imagi k r \cos\vartheta}\frac{\sin(\sqrt{k^2+m^2}t)}{\sqrt{k^2+m^2}}\,,
\end{equation}
where $\Theta(t)$ is the Heaviside theta function, $\Theta(t>0)=1$ and $\Theta(t<0)=0$. The integrals over $\varphi$ yields a factor of $2\pi$ and the one over $\d\cos\vartheta$ a sine function, after which we transform the product of two sines into two cosines, replace $k\to-k$ in the second one, and obtain
\begin{equation}\label{eq:Four4c}
\begin{aligned}
 G_4 &= \frac{\Theta(t)}{4\pi^2 r}\int_{-\infty}^\infty\frac{\d k k}{2\sqrt{k^2+m^2}}\cos\left(\sqrt{k^2+m^2}t-kr\right)\\
 &= -\frac{\Theta(t)}{4\pi^2 r}\frac{\partial}{\partial r}\int_{-\infty}^\infty
 \frac{\d k}{\sqrt{k^2+m^2}}\sin\left(\sqrt{k^2+m^2}t-k r\right)\,.
\end{aligned}
\end{equation}

The last integral in Eq.\ (\ref{eq:Four4c}) can be evaluated with a change of variable for $t>r$ and another one for $r>t$. In both cases, we change the variable $k$ as $k=m\sinh \kappa$, so that $\sqrt{k^2+m^2} = m\cosh \kappa$ and $\d k = m\cosh \kappa \d\kappa$.

In the case $t>r$, we also transform as $t=u\cosh\tau$, $r=u\sinh\tau$, $u=\sqrt{t^2-r^2}$ (invariant interval). This way,
\begin{equation}\label{eq:intT}
 I=\int_{-\infty}^\infty
 \frac{\d k}{\sqrt{k^2+m^2}}\sin\left(\sqrt{k^2+m^2}t-k r\right) = \int_{-\infty}^\infty \d\kappa \sin[mu\cosh(\kappa-\tau)] = \pi J_0(mu)\,,
\end{equation}
where $J_0$ denotes the regular Bessel function of order 0
 (see Refs.\ \cite{RG, Prudnikov1}).

In the case $r>t$, we transform as $t=u\sinh\tau$ and $r=u\cosh\tau$, obtaining
\begin{equation}\label{eq:intS}
 I=\frac{\d k}{\sqrt{k^2+m^2}}\sin\left(\sqrt{k^2+m^2}t-k r\right) = \int_{-\infty}^\infty \d\kappa \sin[mu\sinh(\kappa-\tau)] = 0\,,
\end{equation}
as the integrand is odd.

The results of the integration can be summarised as
\begin{equation}\label{eq:intTS}
 I=\Theta(t-s)\pi J_0(mu)\,.
\end{equation}
Upon derivation w.r.t.\ $r$, we obtain
\begin{equation}\label{eq:KGG}
 G({\bf x},t) = \frac{\delta(t-r)}{4\pi r} -\frac{m\Theta(t-r)}{4\pi\sqrt{t^2-r^2}}J_1\left(m\sqrt{t^2-r^2}\right)\,.
\end{equation}
In Ref.\ \cite{IS}, the calculation is performed for arbitrary (integer) dimension of spacetime.

A good consistency check of our calculation is now comparing the Green function of the time dependent Klein-Gordon equation, and that of $\nabla^2-m^2$ obtained in Sec.\ \ref{ssec:lin}. What one needs to show is that
\begin{equation}\label{eq:consistency}
 \int\d t G_4({\bf x},t) = G({\bf x})\,.
\end{equation}
The Dirac delta term yields $1/(4\pi r)$. In the second term, due to the $\Theta$ function, we need to integrate over $t$ from $r$ to infinity. By a change of variable $r=rt'$, we are led to the integral (Refs.\ \cite{RG, Prudnikov2})
\begin{equation}\label{eq:inCons}
 \int_1^\infty \frac{J_1(b\sqrt{x^2-1})}{\sqrt{x^2-1}}\d x = \frac{1}{b}(1-\e^{-b})\,.
\end{equation}

\section{Interaction potential in the Klein-Gordon model}\label{app:KGInt}
Let us consider interaction energy in the massive Klein-Gordon model, i.e., a massive real scalar field, coupled to external Dirac-delta sources. The field satisfies the equation (for the static case)
\begin{equation}\label{eq:KGS}
 (\nabla^2-m_s^2)\phi = \sigma = s_1 \delta^{(3)}({\bf x}) + s_2 \delta^{(3)}({\bf x}-{\bf x}_0)\,.
\end{equation}
Using the linearity of the equation, we split the field as $\phi=\phi_1+\phi_2$, both satisfying
\begin{equation}\label{eq:KGSsplit}
 (\nabla^2-m_s^2)\phi_1 = \sigma_1 = s_1 \delta^{(3)}({\bf x})\,,\quad
 (\nabla^2-m_s^2)\phi_2 = \sigma_2 = s_2 \delta^{(3)}({\bf x}-{\bf x}_0)\,.
\end{equation}
The energy density of a static Klein-Gordon field is
\begin{equation}\label{eq:KGErg}
\mathcal{E}_{\rm KG}=\frac{1}{2}|\nabla\phi|^2 + \frac{1}{2}m_s^2\phi^2+\sigma\phi\,.
\end{equation}
The interaction energy is therefore the cross term in the energy of $\phi_1+\phi_2$
\begin{equation}\label{eq:KGErgInt}
\mathcal{E}_{\rm KG, int}=\nabla\phi_1\nabla\phi_2 + m_s^2 \phi_1 \phi_2 + \sigma_1\phi_2+\sigma_2\phi_2\,,
\end{equation}
and the interaction potential between the two particles corresponding to the Dirac delta sources is
\begin{equation}\label{eq:IntPotKG}
V_{\rm KG}=\int \d^3 x \mathcal{E}_{\rm KG, int}\,,
\end{equation}
where, in the first term, we apply the identity $\nabla\phi_1 \nabla\phi_2 = \nabla(\phi_1\nabla\phi_2) - \phi_1\nabla^2\phi_2$. Here, upon integration, the first term gives a vanishing contribution,
\begin{equation}\label{eq:GaussTrick}
 \int\d^3 x \nabla(\phi_1\nabla\phi_2) = \int \phi_1\nabla\phi_2 \d^2{\bf s}=0\,,
\end{equation}
where the second interation is taken over a large sphere of radius $R$, in the limit $R\to\infty$, and $\d^2{\bf s}$ is the surface element vector; the fields vanish exponentially, therefore, the integral vanishes. What remains, cancels $\phi_1\sigma_2$, yielding
\begin{equation}\label{eq:IntPotKGv}
V_{\rm KG} = \int \d^3 x \phi_2 \sigma_1 = s_1 \phi_2(0)\,,
\end{equation}
using Eq.\ (\ref{eq:KGSsplit}) the properties of the Dirac delta. Now, using the Green function of the operator $\nabla^2 - m_2^2$, $G=-1/(4\pi r)\exp(-m_s r)$, we obtain
\begin{equation}\label{eq:IntPotKGFinal}
 V_{\rm KG} = -\frac{s_1 s_2}{4\pi r}\e^{-m_s r}\,,\quad r=|{\bf x}_0|\,.
\end{equation}
Note, that although we have used the field equation for $\phi_2$, the result is symmetric. Had we used the vector identity as $\nabla\phi_1 \nabla\phi_2 = \nabla(\nabla\phi_1\phi_2) - \nabla^2\phi_1 \phi_2$, and the field equation for $\phi_1$ from Eq.\ (\ref{eq:KGSsplit}), we would have arrived at the same result.

\section{Interaction potential in a Proca field}\label{app:PInt}
A static, purely electric Proca field with two point sources, one at zero, satisfy the field equation
\begin{equation}\label{eq:Procasrc}
 (\nabla^2-m_A^2)A_0 = -\rho = - q_1 \delta^{(3)}({\bf x}) - q_2 \delta^{(3)}({\bf x}-{\bf x}_0)\,,
\end{equation}
and the energy of the static, purely electric Proca field is
\begin{equation}\label{eq:ProcaErg}
 \mathcal{E}_{\rm P} = -\frac{1}{2}(\nabla A_0)^2 - \frac{1}{2}m^2 (A_0)^2 +\rho A_0\,.
\end{equation}
We shall apply a similar splitting procedure as in the case of the Klein-Gordon field, $A_0=A_{1,0}+A_{2,0}$,
\begin{equation}\label{eq:ProcaSplit}
 (\nabla^2-m_A^2)A_{1,0} = -\rho_1 = q_1 \delta^{(3)}({\bf x})\,,\quad
 (\nabla^2-m_A^2)A_{2,0} = -\rho_2 = q_2 \delta^{(3)}({\bf x}-{\bf x}_0)\,.
\end{equation}
The interaction energy, as before, is
\begin{equation}\label{eq:PErgInt}
 \mathcal{E}_{\rm P,int} = -\nabla A_{1,0} \nabla A_{2,0} -m^2 A_{1,0}A_{2,0} +\rho_1 A_{2,0} + \rho_2 A_{1,0}\,,
\end{equation}
and with the same partial integration technique,
using now the Green function of $-\nabla^2+m^2$, $G=1/(4\pi r)\e^{-m_A r}$,
\begin{equation}\label{eq:IntPotP}
 V_{\rm P} = \frac{q_1 q_2}{4\pi r}\e^{-m_A r}
\end{equation}
is obtained.

\def\refttl#1{{\sl ``#1''}, }%

\end{document}